\documentclass[pra,twocolumn,superscriptaddress,amsmath,amssymb,floatfi,noshowpacs]{revtex4}
\usepackage{graphicx}

\begin{document}
\title{\textbf{Casimir forces in transmission-line circuits: QED and fluctuation-dissipation formalisms}}
%\date{\today}
\author{Ephraim Shahmoon}
\affiliation{Department of Physics, Harvard University, Cambridge MA 02138, USA}
\date{\today}

\begin{abstract}
Transmission-line waveguides can mediate long-range fluctuation-induced forces between neutral objects.
We present two approaches for the description of these forces between electric components embedded in transmission-line circuits. The first, following ordinary quantum electrodynamics (QED), consists of the quantization and scattering theory of voltage and current waves inside transmission lines. The second approach relies on a simple circuit analysis with additional noisy current sources due to resistors in the circuit, as per the fluctuation-dissipation theorem (FDT). We apply the latter approach to derive a general formula for the Casimir force induced by circuit fluctuations between any two impedances. The application of this formula, considering the sign of the resulting force, is discussed. While both QED and FDT approaches are equivalent, we conclude that the latter is simpler to generalize and solve.
\end{abstract}

 \maketitle
\section{Introduction}
The ongoing experimental progress in the ability to efficiently couple polarizable emitters to waveguide structures opens many exciting possibilities in quantum optics research \cite{RAU1,RAU3,KIM2,KIM3,LOD,ALP,WAL1,WAL2}. A central theme in this growing field is the possibility to design long-range and modified interactions between polarizable objects, mediated by the waveguide modes. Such enhanced interactions should prove useful for entanglement generation \cite{GON,RDDI} and for the study of novel many-body phenomena and quantum nonlinear optics with long-range interactions \cite{CHA1,RIT,LIDDIna,CHA2,GOR,RAB,EITNLO,CHA3}.

Quantum vacuum fluctuations of waveguide modes may also drive and mediate novel Casimir and van der Waals forces, with drastically modified distance dependencies \cite{vdWTL,MAZ,SIL,vdWMWG,HAK,FAR,SCH}. Of particular interest are such vacuum-induced interactions mediated by transmission-line waveguides, which can become extremely long-range and strong \cite{vdWTL,MAZ,SIL}, as opposed to e.g. exponentially decreasing interactions mediated by other waveguides \cite{vdWMWG,HAK,FAR}.
The unique feature of electromagnetic phenomena in transmission-line (TL) waveguides is that it can be described by electrodynamics in one-dimension (1d) , or equivalently, by TL circuit theory. This is in contrast to electrodynamics in other types of waveguides, for which a 1d description of broadband and non-resonant phenomena, such as zero-point fluctuation forces, is not adequate. Transmission-line circuits in the quantum regime are now routinely used in the field of circuit quantum electrodynamics, where squeezed vacuum generation and the dynamical Casimir effect were recently observed \cite{WAL3,SID,WIL,PARA}.

The aim of the present work is to consider a general theoretical framework for the description of quantum fluctuation-induced potentials inside TL circuits. As explained below, we focus on the regime at which the 1d electrodynamic description is valid, and compare two equivalent approaches for such a theoretical framework. The first approach considered here is a straightforward adaptation of QED to TL circuits, consisting of canonical quantization of propagating voltage and current fields \cite{YUR,DEV}. Quantum fluctuations and associated phenomena then originate in those quantized fields, and are described by an effective 1d QED theory, combined with a 1d scattering formalism \cite{vdWTL,REY}.

In an alternative approach, quantization is performed only implicitly, by using the fluctuation-dissipation theorem (FDT) \cite{LL}. Here, quantum fluctuations in the form of noisy current sources, emerge from resistive elements within the circuit, in analogy to the Lifshitz treatment of the Casimir force between dispersive materials in free space \cite{DZO,BAR,MIL,FoQVs}. The inclusion of these noisy currents within standard circuit theory, then provides an elegant route towards the calculation of fluctuation potentials in circuits, by simply using the Kirchhoff circuit laws. Moreover, much like in the Lifshitz treatment, the advantage of the FDT description over that using canonical quantization, is the possibility to treat dissipative systems.

The paper is organized as  follows. In Sec. II we briefly review a few essential elements of transmission-line theory and present the Casimir problem adapted to circuits. Sec. III is devoted to the introduction of the QED and FDT approaches and the identification of the driving quantum fluctuations sources therein. We proceed to the calculation of the Casimir force between two circuit components within the QED and FDT approaches in Sec. IV and V, respectively, where in the latter the calculation is generalized to dissipative components and terminated lines. As an illustration, we discuss in Sec. VI a few simple examples, where special care is taken to the possibility of repulsive forces. We conclude by a discussion of a few formal aspects and possible experimentally relevant systems in Sec. VII.

\section{Transmission-line circuits}
\subsection{The TEM mode: 1d electrodynamics}
A transmission line (TL) is formed by at least two conductors (Fig. 1a). This guarantees the existence of the transverse-electromagnetic (TEM) mode as the fundamental transverse mode of the TL \cite{KONG}. The TEM mode function, $E_0 e^{i k x}$, where $x$ is the propagation direction and $E_0$ is independent of frequency, along with its linear dispersion, $\omega=|k| c$, with $k$ the wavenumber along $x$ and $c$ the speed of propagation, constitute 1d wave propagation at \emph{all} frequencies. This unique 1d propagation is in contrast to the guided modes of dielectric waveguides (such as optical fibers) or hollow metallic waveguides, where $E_0$ is frequency dependent \cite{KONG}. All other guided modes supported by the TL possess a cutoff frequency which is higher than $c/a$, $a$ being the separation between the conductors that form the TL (Fig. 1a). Therefore, the 1d description of electromagnetic wave propagation in TL circuits is valid for frequencies $\omega<c/a$, where only the TEM mode exists. For quantum fluctuation forces, where the driving fluctuations may exist in an immense frequency bandwidth, an electrodynamic description based only on the TEM mode is valid for $l>a$, $l$ being the distance between the interacting objects \cite{vdWTL}. In principle, the TEM mode possesses  two orthogonal field polarizations, however, due to the anisotropy of circuit elements (e.g. parallel-plate capacitors), we consider only a single polarization, resulting in the usual scalar TL description.

\begin{figure}
\begin{center}
\includegraphics[width=\columnwidth]{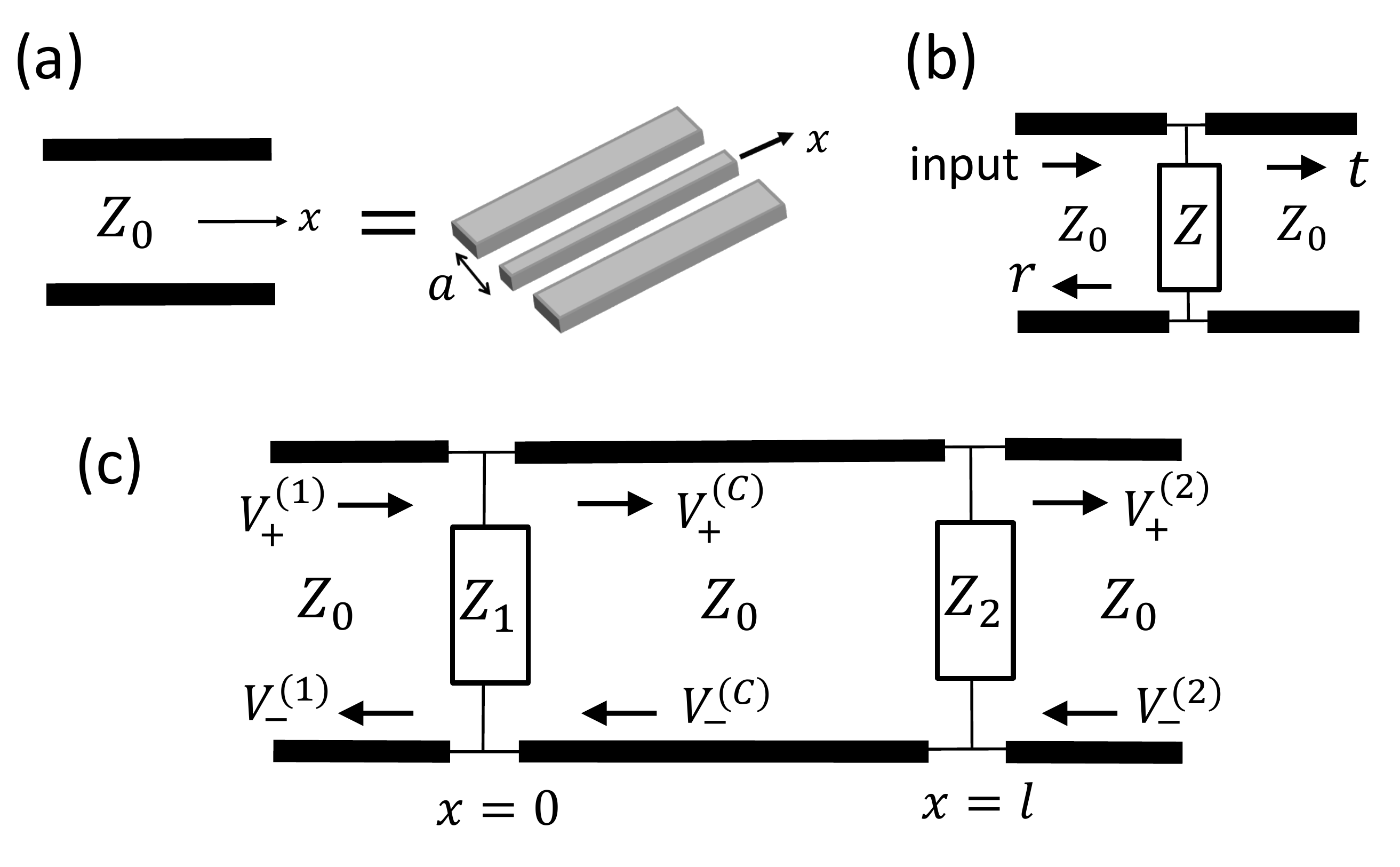}
\caption{\small{
Transmission-line (TL) circuits and the Casimir force. (a) A TL (two parallel thick lines) is characterized by the impedance $Z_0$ of its fundamental TEM mode. Practical realizations range from a pair of parallel wires to two separated conductors in a planar geometry, such as in the widely used coplanar waveguide (see right-hand side). (b) A mirror formed by an impedance $Z$ embedded in a TL: incident waves are scattered forward ($t$) and backward ($r$) (Eq. \ref{rt}). (c) A straightforward adaptation of the Casimir problem to TL circuits. Quantum fluctuations of the quantized TEM mode supported by the TL are scattered off the two circuit elements with impedances $Z_1$ and $Z_2$, resulting in an interaction and force between the elements.
 }} \label{fig1}
\end{center}
\end{figure}

\subsection{Voltage and current waves}
Considering only the TEM mode, the electromagnetic field can be described by propagating voltage and current waves in 1d, obeying the so-called telegraphers equations \cite{POZ}
\begin{equation}
\partial_x V=-L'\dot{I}, \quad \partial_x I=-C'\dot{V},
\label{TL}
\end{equation}
with $V(x,t)$ and $I(x,t)$ the voltage and current waves at time $t$ and position $x$ along the TL. For a given section of the TL with capacitance and inductance per unit length, $C'$ and $L'$, respectively, that are independent of $x$, this leads to the 1d Helmholtz equation in this section,
\begin{equation}
(\partial_x^2+k^2)u, \quad k=\omega/c, \quad c=1/\sqrt{L'C'},
\label{HELM}
\end{equation}
for $u=V,I$ written in frequency domain. The resulting forward and backward propagating voltage waves $V_{\pm k} e^{\pm ik x}$ can then be reflected by an impedance $Z(\omega)$ that terminates the line, with a reflection coefficient \cite{POZ}
\begin{equation}
r(\omega)=\frac{Z(\omega)-Z_0}{Z(\omega)+Z_0}, \quad Z_0=\sqrt{L'/C'},
\label{r}
\end{equation}
and where $Z_0$ is the line impedance (real number). In turn, the forward/backward current amplitudes are related to their respective voltage amplitudes by $I_{\pm k}=\pm V_{\pm k}/Z_0$. When the impedance $Z$ is placed in between two semi-infinite TL sections (Fig. 1b), it acts as a mirror with reflection and transmission coefficients
\begin{equation}
r=\frac{Z||Z_0-Z_0}{Z||Z_0+Z_0}, \quad t=1+r, \quad Z||Z_0\equiv \frac{Z  Z_0}{Z+Z_0}.
\label{rt}
\end{equation}
Here we used the fact that the load of a semi-infinite TL seen by the circuit is its characteristic impedance $Z_0$ (connected in parallel to $Z$). We note that $Z$ may in general contain dissipative elements (a real part), in which case $|r|^2+|t|^2\neq 1$ (see Appendix A).

\subsection{Force between circuit components}
The electromagnetic energy of an infinitesimal section $dx$ at position $x$ of the TL is given by that stored in its capacitance and inductance.  The electromagnetic energy $H$ of the TL is then
\begin{equation}
H=\int_{-\infty}^{\infty}dx H'(x), \quad H'(x)=\frac{1}{2}L'(x)I^2(x)+\frac{1}{2}C'(x)V^2(x),
\label{H}
\end{equation}
with $H'(x)$ being the energy density (energy per unit length). Considering two lumped-element components with impedances $Z_1$ and $Z_2$ at positions $x=0$ and $x=l$ along the line (Fig. 1c), the force between the components mediated by the TL's TEM modes is given by the discontinuity of the electromagnetic energy density around each component (see Appendix B)
\begin{equation}
f=H'(0^-)-H'(0^+)=H'(l^+)-H'(l^-).
\label{f}
\end{equation}
Here, the positive sign ($f>0$) means an attractive force between the components. In the case of a Casimir-like force, the task is therefore to find the voltage/current fields around the components (e.g. at $x=0^+$ and $x=0^-$), given the sources of quantum fluctuations that exist in the circuit, and insert them in Eqs. (\ref{H}) and (\ref{f}).

\section{Source of quantum fluctuations}

\subsection{QED approach}
Within QED, the Casimir interaction is driven by the vacuum fluctuations of the electromagnetic field modes that are coupled to the interacting objects. Therefore, the first step is to perform a canonical quantization of the field modes of a "free" TL (where $C'$ and $L'$ are independent of $x$). This has been done before by many authors \cite{YUR,DEV,BLA,HEN} and is also presented in Appendix C. In short, we begin by writing the 1d Helmholtz equation (\ref{HELM}) for the charge wave $Q(x,t)=\int dt I(x,t)$, find the Lagrangian that yields it, and the canonical conjugate to $Q(x)$, $\phi(x)=L'I(x)$. Then, moving to Hamiltonian formulation, quantization is performed by imposing the commutation relations $[\hat{Q}(x), \hat{\phi}(x')]=i\hbar\delta(x-x')$. The resulting quantized voltage and current fields take the form (Appendix C)
\begin{eqnarray}
\hat{V}(x)&=&\sum_{k>0}\left[\left(\hat{V}_{+k}e^{ikx}+\hat{V}_{-k}e^{-ikx}\right)+\mathrm{h.c.}\right],
\nonumber \\
\hat{I}(x)&=&\frac{1}{Z_0}\sum_{k>0}\left[\left(\hat{V}_{+k}e^{ikx}-\hat{V}_{-k}e^{-ikx}\right)+\mathrm{h.c.}\right].
\label{VI}
\end{eqnarray}
Here, the terms in parentheses are the so-called positive-frequency components of the fields, and the forward/backward ($+/-$) amplitudes of waves at frequency $\omega=kc$ are given by
\begin{equation}
\hat{V}_{\pm k}=\mp i\sqrt{\frac{\hbar c k}{2 C' L}}\hat{a}_{\pm k},
\label{a}
\end{equation}
where $L$ is a quantization length of the 1d plane wave modes $e^{ikx}/\sqrt{L}$, and where $\hat{a}_{\mu k}$ ($\mu=\pm$) possess the commutation relations $[\hat{a}_{\mu k},\hat{a}^{\dag}_{\mu'k'}]=\delta_{\mu\mu'}\delta_{k k'}$, $[\hat{a}_{\mu k},\hat{a}_{\mu'k'}]=0$.  The Hamiltonian operator is given by Eq. (\ref{H}) with $V(x)$ and $I(x)$ now replaced by the operators from Eq. (\ref{VI}), or, in terms of the operators $\hat{a}_{\pm k}$ by $H=\sum_{k>0}\hbar c k (\hat{a}^{\dag}_{+k}\hat{a}{+k}+\hat{a}^{\dag}_{-k}\hat{a}{-k}+1)$.

In summary within the QED formulation, the source of Casimir forces are the quantum fluctuations of the voltage and current fields (\ref{VI}), which propagate inside the TL with the amplitudes (\ref{a}). The way these fields propagate and scatter between different impedances determines the energy density and the force in Eq. (\ref{f}).

\subsection{FDT approach}
According to the FDT, the dissipation of electromagnetic energy in a resistor (ohmic losses) must be accompanied by current fluctuations $I_N(t)$ inside it. At open circuit, when the resistor is disconnected, the spectrum of these fluctuations is given by (Appendix D) \cite{DEV,NYQ}
\begin{equation}
S_{I_N}(\omega)=\frac{1}{T_e}\langle I_N(\omega) I_N(-\omega)\rangle=\frac{2\hbar\omega}{R(\omega)}\frac{1}{1-e^{-\beta \hbar \omega}},
\label{IN}
\end{equation}
with a corresponding voltage spectrum of $S_{V_N}=R^2 S_{I_N}$, and where $I_N(\omega)=\int_{-\infty}^{\infty}dt e^{i\omega t}I_N(t)$, $T_e\rightarrow\int_{-\infty}^{\infty}dt=2\pi\delta(\omega=0)$ is the duration of the experiment, and $\beta$ is the inverse temperature. These fluctuations originate from the internal degrees of freedom of the resistor and hence should exist regardless of whether the resistor is connected to other components or not. Therefore, in order to account for these fluctuations within circuit theory, the resistor model is accompanied by a shunt current source with the spectrum (\ref{IN}) (Fig. 2a; noting also the equivalence to a voltage source in series \cite{HEN}). This guarantees the constant supply of the current noise as per the FDT.

\begin{figure}
\begin{center}
\includegraphics[scale=0.3]{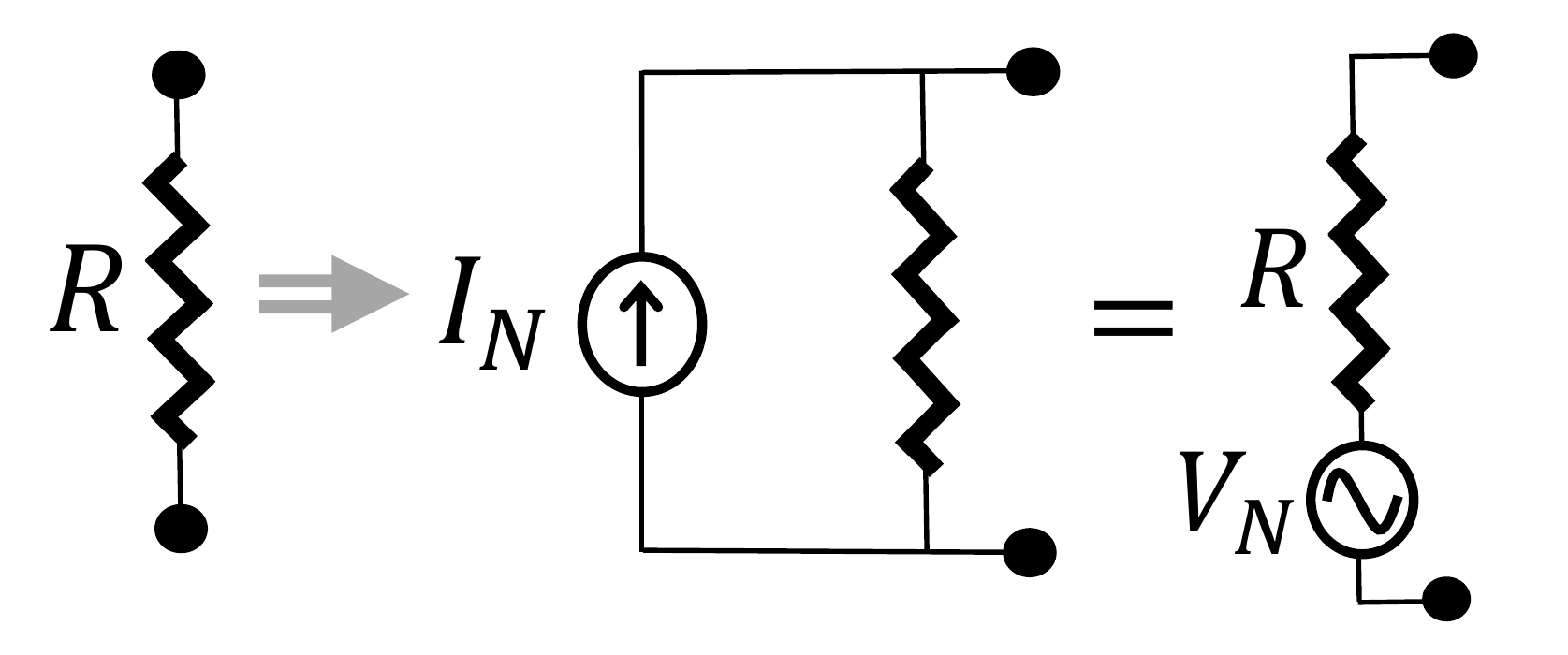}
\caption{\small{
Source of quantum fluctuations within the fluctuation-dissipation theorem (FDT) formalism. Quantum noise emerges from resistive elements, which are modelled by their resistance $R$ connected in parallel with a current source $I_N$ with the spectrum of Eq. (\ref{IN}) (equivalently, in series with a voltage $V_N=R I_N$).
 }} \label{fig2}
\end{center}
\end{figure}

To conclude, the FDT approach for the calculation of quantum fluctuation forces in circuits goes as follows. In a given circuit, all resistors (or equivalently, semi-infinite TLs) are modeled by their resistance $R$ in parallel to a current noise source with the spectrum (\ref{IN}). By applying the Kirchhoff laws in frequency domain, one can then find voltages and currents in the circuit and hence the energy density and force (\ref{f}).

\subsection{Equivalence between both approaches}
As already mentioned above, the impedance of a semi-infinite TL seen by a circuit (e.g. an electrical component), is equivalent to a resistance $R=Z_0$, with $Z_0$ the characteristic impedance of the TL. This is a manifestation of the fact that an infinite space, here 1d space, does not reflect incident waves, and is hence equivalent to dissipation, which is represented by a resistor in the case of circuits. Following the reasoning of the FDT, we would then expect that the noise inserted into a circuit from a semi-infinite TL with a characteristic impedance $R$, is identical to that inserted by a resistor $R$ (Fig. 3). This fluctuation-source equivalence can be demonstrated as follows.

First, consider the spectrum of voltage fluctuations inserted by a resistor $R$ into a circuit. The input of any circuit is represented by its wires, characterized by the impedance $Z_0$ of a semi-infinite TL (Fig. 3a). The task is then to find the voltage fluctuations induced by $R$ that enter and propagate into this TL at $x=0$. The forward-propagating current $I_+(0)$ at $x=0$ is easily found in frequency domain in terms of the current source $I_N$, by effectively substituting the TL with an impedance $Z_0$ and using Kirchhoff's laws, obtaining $I_+(0)=I_N R/(R+Z_0)$. The forward-propagating voltage is given by $V_+(0)=Z_0 I_+(0)$ (see Sec. IIB), yielding the input fluctuation spectrum
\begin{equation}
S_{V_+(0)}(\omega)=\frac{\langle V_+(0,\omega) V_+(0,-\omega)\rangle}{T_e}=\frac{R Z_0^2}{(R+Z_0)^2}\frac{2\hbar\omega}{1-e^{-\beta\hbar\omega}},
\label{SFDT}
\end{equation}
where the FDT fluctuation source of Eq. (\ref{IN}) was used.

\begin{figure}
\begin{center}
\includegraphics[width=\columnwidth]{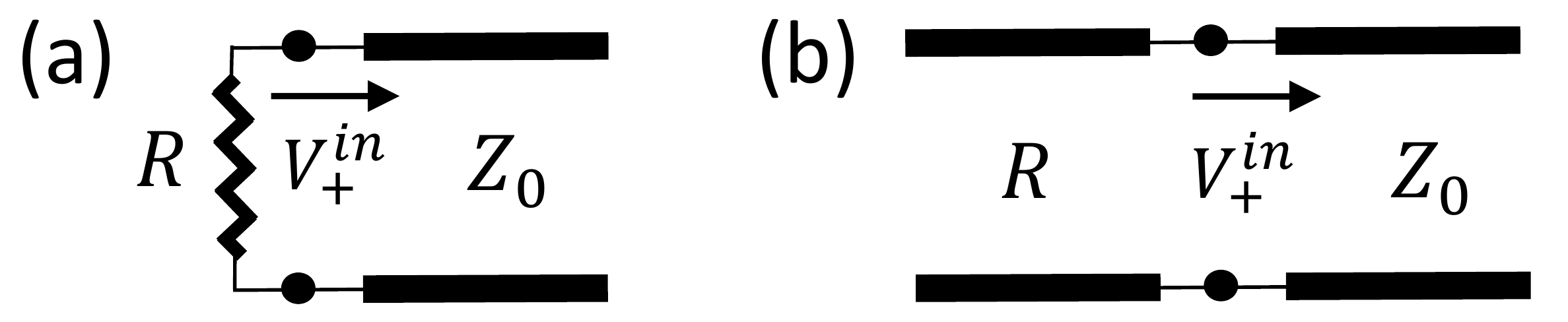}
\caption{\small{
Equivalence between the QED and FDT approaches: Noise input. The voltage fluctuations inserted by a resistor $R$ into a TL with impedance $Z_0$ [case (a), calculated using the FDT approach] is identical to those inserted by a semi-infinite TL with impedance $R$ [case (b), calculated using the QED approach].
 }} \label{fig3}
\end{center}
\end{figure}

In analogy, consider now a similar problem, where the resistor $R$ is replaced by a semi-infinite TL with characteristic impedance $R$ (Fig. 3b). The forward-propagating voltage in the TL $Z_0$ at $x=0^+$, inserted by the TL $R$, is given by $V_+(0^+)=t_+V_+(0^-)$ with fluctuation spectrum
\begin{equation}
S_{V_+(0^+)}(\omega)=|t_+|^2S_{V_+(0^-)}(\omega), \quad  t_+=2Z_0/(Z_0+R).
\label{SQED}
\end{equation}
 Here $V_+(0^-)$ is the forward-propagating voltage fluctuation in the TL $R$ and $t_+=1+r_+$ is the transmission coefficient from left to right around $x=0$ [Eq. (\ref{r}) for $Z(\omega)=R$]. The remaining task is to find $S_{V_+(0^-)}(\omega)$, which is the spectrum of the freely forward propagating fluctuations in a TL with characteristic impedance $R\equiv\sqrt{L'/C'}$. This can be done by considering the voltage operator of a free TL in the QED formulation, Eqs. (\ref{VI}) and (\ref{a}). In the Heisenberg picture with respect to Hamiltonian (\ref{H}), the operators evolve in time as $\hat{a}_{\pm k}(t)=\hat{a}_{\pm k}e^{-i ck t}$ so that
\begin{equation}
\hat{V}_+(x,t)=\sum_{k>0}\sqrt{\frac{\hbar c k}{2C'L}}\left[-i\hat{a}_{+k}e^{ik(x-ct)}+i\hat{a}^{\dag}_{+k}e^{-ik(x-ct)}\right].
\label{Vp}
\end{equation}
Assuming a thermal state for the voltage modes in TL $R$ at $x<0$, we take
\begin{equation}
\langle \hat{a}_{+k}^{\dag}\hat{a}_{+k'}\rangle=\delta_{kk'}n_k, \quad \langle \hat{a}_{+k}\hat{a}_{+k'}^{\dag}\rangle=\delta_{kk'}(n_k+1),
\label{nk}
\end{equation}
with $n_k=n(\omega)=1/(e^{\beta \hbar \omega}-1)$ for $\omega=ck$. Inserting Eq. (\ref{nk}) into Eq. (\ref{Vp}) in the continuum limit $\sum_{k>0}\rightarrow (L/2\pi c)\int_0^{\infty}d\omega$, the voltage correlation at $x=0^-$ is obtained as
\begin{eqnarray}
&&\langle\hat{V}_+(0^-,t)\hat{V}_+(0^-,0)\rangle
\nonumber\\
&&=\int_0^{\infty}\frac{d\omega}{2\pi}\frac{\hbar \omega}{2C' c}\left[(n(\omega)+1)e^{-i\omega t}+n(\omega)e^{i\omega t}\right].
\label{Rv}
\end{eqnarray}
By definition, the first term of the integrand in Eq. (\ref{Rv}) is $S_{V_+(0^-)}(\omega)$, which is inserted into Eq. (\ref{SQED}) with $R=\sqrt{L'/C'}$, to yield exactly the same spectrum as in Eq. (\ref{SFDT}).

\section{Casimir force between circuit components: QED formulation}
Reconsidering the configuration from Fig. 1c, we shall now turn to the calculation of the force between the electric components, Eq. (\ref{f}). The QED formalism is particulary useful for non-dissipative components for which the impedances $Z_{1,2}$ are purely imaginary. These components then form non-absorptive scatterers, or mirrors, with the reflection and transmission coefficients, Eqs. (\ref{rt}), satisfying $|r|^2+|t|^2=1$ (Appendix A). The expression for the force in Eq. (\ref{f}) is given in terms of the Hamiltonian density, which in turn requires the knowledge of the voltage and current fields around the scatterers. These are found below by a 1d scattering formulation.

\subsection{The scattering problem}
Since scattering problems are conveniently treated in frequency domain, we first write the Hamiltonian density $H'(x)$ from Eq. (\ref{H}) using the voltage and current operators from Eq. (\ref{VI}). Taking the quantum mechanical average over $H'(x)$, we assume a thermal state for the input field fluctuations arriving from both semi-infinite ends of the TL, so that $\langle \hat{V}_{\mu k} \hat{V}_{ \mu' k'}\rangle=0$,  $\langle \hat{V}^{\dag}_{ \mu k} \hat{V}_{\mu' k'}\rangle,\langle \hat{V}_{\mu k}\hat{V}^{\dag}_{ \mu' k'}\rangle\propto \delta_{k k'}$ ($\mu,\mu'\in\{\pm\}$). The Hamiltonian density then assumes the following spectral representation
\begin{equation}
H'=C'\sum_{k>0}\langle\hat{V}_{+k}\hat{V}^{\dag}_{+k}+\hat{V}_{-k}\hat{V}^{\dag}_{-k}+\hat{V}^{\dag}_{+k}\hat{V}_{+k}+\hat{V}^{\dag}_{-k}\hat{V}_{-k}\rangle,
\label{HQED}
\end{equation}
which depends on $x$ only through the TL section wherein $x$ is located. Therefore, for the force (\ref{f}), we need to find the amplitudes $\hat{V}_{\pm k}$ of the forward and backward propagating voltages at the different TL sections, e.g. those of $x<0$ and $0<x<l$ around the scatterer $Z_1$ at $x=0$.

This defines the following scattering problem (as in Ref. \cite{REY}). We divide the 1d space into three sections 1,2 and C for $x<0$, $x>l$ and $0<x<l$, respectively. Then, the field amplitudes of all sections are found as a function of input fields, consisting of the freely-propagating amplitudes, $\hat{V}^{(1)}_{+k}=\hat{V}_{+k}$ (forward-propagating, section 1) and  $\hat{V}^{(2)}_{-k}=\hat{V}_{-k}$ (backward-propagating, section 2), with $\hat{V}_{ \pm k}$ from Eq. (\ref{a}).
The solution is performed as usual by matching the fields at the boundaries $x=0$ and $x=l$, obtaining
\begin{eqnarray}
&&\hat{V}^{(C)}_{+ k}=t_1 \hat{V}_{+ k}+r_1 \hat{V}^{(C)}_{- k},
\nonumber \\
&&\hat{V}^{(C)}_{- k}e^{-ikl}=r_2 \hat{V}^{(C)}_{+ k}e^{ikl}+t_2 \hat{V}_{- k}e^{-ikl},
\label{bc}
\end{eqnarray}
with $r_{1,2}$ and $t_{1,2}$ the reflection/transmission coefficients, Eqs. (\ref{rt}), for purely imaginary impedances $Z=Z_{1,2}$. Solving for the field amplitudes in terms of the inputs, $\hat{V}_{\pm k}$, we find
\begin{equation}
\left(
  \begin{array}{c}
    \hat{V}^{(C)}_{+k} \\
    \hat{V}^{(C)}_{-k} \\
  \end{array}
\right)
  =\overline{\overline{R}}
\left(
  \begin{array}{c}
    \hat{V}_{+k} \\
    \hat{V}_{-k} \\
  \end{array}
\right),
\quad
\left(
  \begin{array}{c}
    \hat{V}^{(2)}_{+k} \\
    \hat{V}^{(1)}_{-k} \\
  \end{array}
\right)
  =\overline{\overline{S}}
\left(
  \begin{array}{c}
    \hat{V}_{+k} \\
    \hat{V}_{-k} \\
  \end{array}
\right),
\label{sQED}
\end{equation}
with the reflection and scattering matrices,
\begin{eqnarray}
\overline{\overline{R}}&=&\frac{1}{1-r_1r_2e^{i2kl}}\left(
                          \begin{array}{cc}
                            t_1 & r_1t_2 \\
                            r_2t_1e^{i2kl} & t_2 \\
                          \end{array}
                        \right),
\nonumber\\
\overline{\overline{S}}&=&\left(
                          \begin{array}{cc}
                            \frac{t_1t_2}{1-r_1r_2e^{i2kl}} & r_2e^{-i2kl}+\frac{r_1t_2^2}{1-r_1r_2e^{i2kl}} \\
                           r_1+\frac{r_2t_1^2e^{i2kl}}{1-r_1r_2e^{i2kl}}   & \frac{t_1t_2}{1-r_1r_2e^{i2kl}} \\
                          \end{array}
                           \right).
\label{RS}
\end{eqnarray}

\subsection{The force}
The force, calculated on the impedance $Z_1$, is given by $f=H'^{(1)}-H'^{(C)}$, with $H'^{(n)}$ the Hamiltonian density in section $n=1,C$. Beginning with $H'^{(1)}$ we insert $\hat{V}^{(1)}_{+k}=\hat{V}_{+k}$ and $\hat{V}^{(1)}_{- k}$ from Eqs. (\ref{sQED}) and (\ref{RS}) into Eq. (\ref{HQED}). Taking the statistics of a thermal state, Eq. (\ref{nk}), for the input fields $\hat{V}_{\pm k}$ from Eq. (\ref{a}),  and the continuum limit $\sum_{k>0}\rightarrow (L/2\pi)\int_0^{\infty}d k$, we find
\begin{equation}
H'^{(1)}=\int_0^{\infty}\frac{dk}{2\pi}\frac{\hbar ck}{\tanh\left(\frac{\beta\hbar ck}{2}\right)}\frac{1+|\rho_1|^2+|\tau|^2}{2},
\label{H1a}
\end{equation}
with the notation $\overline{\overline{S}}=\left(
                                             \begin{array}{cc}
                                               \tau & \rho_2 \\
                                               \rho_1 & \tau \\
                                             \end{array}
                                           \right)$
for the scattering matrix from Eq. (\ref{RS}). For the non-absorptive scatterers case above, where $Z_{1,2}$ are purely imaginary and $|r_{1,2}|^2+|t_{1,2}|^2=1$, we verify that $|\rho_1|^2+|\tau|^2=1$ as it should, such that
\begin{equation}
H'^{(1)}=\int_0^{\infty}\frac{dk}{2\pi}\frac{\hbar ck}{\tanh\left(\frac{\beta\hbar ck}{2}\right)}.
\label{H1}
\end{equation}
For $H^{(C)}$ we insert $\hat{V}^{(C)}_{\pm k}$ from Eqs. (\ref{sQED}) and (\ref{RS}) inside Eq. (\ref{HQED}) and find
\begin{equation}
H'^{(C)}=\int_0^{\infty}\frac{dk}{2\pi}\frac{\hbar ck}{\tanh\left(\frac{\beta\hbar ck}{2}\right)}
\frac{1-|r_1|^2|r_2|^2}{|1-r_1r_2e^{i2kl}|^2},
%\frac{|t_1|^2(1+|r_2|^2)+|t_2|^2(1+|r_1|^2)}{2|1-r_1r_2e^{i2kl}|^2}.
\label{HC}
\end{equation}
where again $|r_{1,2}|^2+|t_{1,2}|^2=1$ was used.
Finally, by subtracting Eq. (\ref{HC}) from Eq. (\ref{H1}) we obtain the force
\begin{equation}
f=\int_0^{\infty}\frac{dk}{2\pi}\frac{-\hbar ck}{\tanh\left(\frac{\beta\hbar ck}{2}\right)}\left[\frac{r_1r_2e^{i2kl}}{1-r_1r_2e^{i2kl}}+\frac{r_1^{\ast}r_2^{\ast}e^{-i2kl}}{1-r_1^{\ast}r_2^{\ast}e^{-i2kl}}\right],
\label{fQED}
\end{equation}
identical to that found for 1d scattering in Ref. \cite{REY}.

An interesting point is that of \emph{renormalization}. Both $H'^{(1)}$ and $H'^{(C)}$ are infinite, however only $H'^{(C)}$ depends on the existence of the scatterers and their separation. Therefore, one may view $H'^{(C)}$ as the "bare" force between the scatterers whereas $H'^{(1)}$ as merely a reference term originating in the free TL. Indeed, $H'^{(1)}$ is identical to the energy density of a free TL with $\pm k$ modes traveling at velocity $c$, each with a zero-point energy $\hbar ck/2$.  As such, it can be also associated with the free-TL value of section $C$ rather than that of section 1. Using such an interpretation, the above renormalization is similar to that used in the Lifshitz approach, where the reference energy subtracted from a diverging stress at a given point is that of the corresponding "free" homogenous system \cite{FoQVs}.

\section{Casimir force between circuit components: FDT formulation}

\subsection{Generalized Casimir problem in circuits}
The Casimir problem presented in Fig. 1c and solved in the previous section presents a direct adaptation of the well-known free-space (3d) problem to its TL counterpart, which is nevertheless not the most natural problem to consider in circuits. In particular, the semi-infinite TL sections on the right and left sides of the scatterers $Z_2$ and $Z_1$, respectively, that were introduced therein in analogy to mirrors placed in free space, can be readily replaced by general complex impedances $Z_n$ ($n=1,2$) that \emph{terminate} the line, as in Fig. 4a.
We can decompose any impedance $Z_n$ into a pair of real and imaginary impedances, $R_n$ and $iX_n$, respectively, connected in parallel (Fig. 4b) and satisfying
\begin{equation}
\mathrm{Re}[Z_n]=X_n\frac{R_nX_n}{R_n^2+X_n^2}, \quad \mathrm{Im}[Z_n]=R_n\frac{R_nX_n}{R_n^2+X_n^2}.
\label{Zn}
\end{equation}
The lossless-mirrors problem of the previous section then corresponds to $R_{1,2}=Z_0$, whereas the general $R_{1,2}$ problem is equivalent, in the scattering formulation, to two lossless mirrors $X_{1,2}$  "sandwiched" between two different "media" with characteristic impedances $R_1$ and $R_2$ and which are separated by a medium with characteristic impedance $Z_0$.

The general Casimir problem of Fig. 4a is however naturally understood without the need of such scattering interpretation: the task is to find the zero-point force induced between two electric components due to the fact that they are connected by a wire (TL). The FDT approach used below, allows to solve this problem using the Kirchhoff circuit laws, naturally applied to circuits in the frequency domain. In the following, it will be therefore useful to write the Hamiltonian density of the TL, Eq. (\ref{H}), in frequency representation
\begin{eqnarray}
H'(x)&=&\int_{-\infty}^{\infty}\frac{d\omega}{2\pi}\left[\frac{1}{2}L'S_{I(x)}(\omega)+\frac{1}{2}C'S_{V(x)}(\omega)\right]
\nonumber\\
&=&\frac{1}{2}L'\int_{-\infty}^{\infty}\frac{d\omega}{2\pi}\left[S_{I_+(x)}(\omega)+S_{I_-(x)}(\omega)\right].
\label{HFDT}
\end{eqnarray}
Here the spectrum $S_A(\omega)$ of a signal $A(t)$ is the same as that defined in Eq. (\ref{IN}), and we recall the forward/backward waves at frequency $\omega=kc$ which are defined by $A(x,\omega)=A_+(x) + A_-(x)$ with $A_{\pm}(x)=A_{\pm}(0)e^{\pm ikx}$ and the relation
$I_{\pm}=\pm V_{\pm}/Z_0$ (see Eq. \ref{HELM}).

\begin{figure}
\begin{center}
\includegraphics[width=\columnwidth]{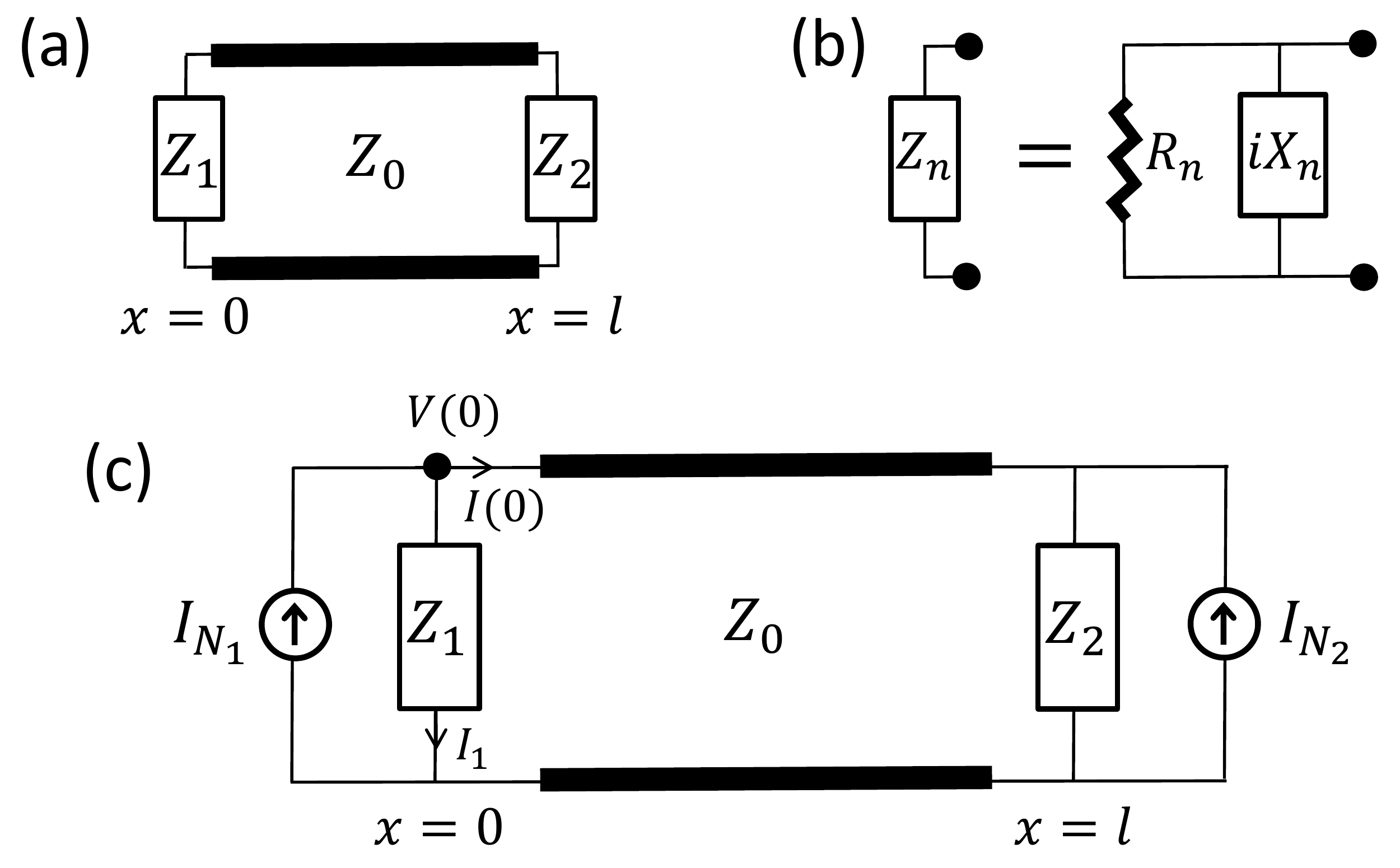}
\caption{\small{
Generalized Casimir problem in circuits. (a) Two generally complex impedances terminate the TL. This more natural formulation does not rely on the existence of semi-infinite TLs at both ends of the impedances. (b) Representation of the impedances $n=1,2$ by parallel dissipative ($R$) and reactive ($X$) parts, Eq. (\ref{Zn}). (c) Circuit diagram of the Casimir problem. The current sources $I_{N_{1,2}}$ are those of Eq. (\ref{IN}) with the resistive parts $R_{1,2}$ from (b). The force is found by a simple solution of the Kirchhoff laws.
 }} \label{fig4}
\end{center}
\end{figure}

\subsection{Circuit solution}
In analogy to Sec. IV A, the idea is to find the Hamiltonian density between the scatterers, e.g. at $x=0^+=0$, from which the force is deduced by a proper renormalization (subtraction). Noting Eq. (\ref{HFDT}), we then need to find the forward and backward current amplitudes at $x=0$, $I_{\pm}(0)$. Using the circuit diagram of Fig. 4b at frequency $\omega$, the current $I_1$ on $Z_1$ is given by
\begin{equation}
I_1=\frac{V(0)}{Z_1}=\frac{1}{Z_1}[V_+(0)+V_-(0)]=\frac{Z_0}{Z_1}[I_+(0)-I_-(0)].
\label{I1}
\end{equation}
Combining it with the Kirchhoff current law at $x=0$, $I_{N_1}=I_1+I(0)$, with $I_{N_1}$ the noise current Eq. (\ref{IN}) with resistance $R_1$ (Fig. 4), we obtain
\begin{equation}
I_{N_1}=I_+(0)\left(1+\frac{Z_0}{Z_1}\right)+I_-(0)\left(1-\frac{Z_0}{Z_1}\right).
\label{K1}
\end{equation}
Similar considerations at $x=l$ lead to
\begin{equation}
I_{N_2}=-I_+(l)\left(1-\frac{Z_0}{Z_2}\right)-I_-(l)\left(1+\frac{Z_0}{Z_1}\right).
\label{K2}
\end{equation}
Considering the definition of the reflection coefficient $r_{n}$ of a line terminated by impedance $Z_{n}$, Eq. (\ref{r}), along with the definition $t_{n}\equiv 1+r_{n}$ for $n=1,2$, the solution of Eqs. (\ref{K1}) and (\ref{K2}) becomes
\begin{eqnarray}
I_+(0)&=&\frac{1}{2(1-r_1r_2e^{i2kl})}\left[t_1I_{N_1}+r_1t_2e^{ikl}I_{N_2}\right],
\nonumber\\
I_-(0)&=&\frac{-e^{ikl}}{2(1-r_1r_2e^{i2kl})}\left[t_2I_{N_2}+ r_2t_1e^{ikl}I_{N_1}\right],
\label{Ipm}
\end{eqnarray}
where the relation between the propagating current waves, $I_{\pm}(l)=I_{\pm}(0)e^{\pm ikl}$, was used.

\subsection{The force}
Assuming that the noise input from the two resistances $R_{1,2}$ (Fig. 4) are uncorrelated, and considering their corresponding noise spectra from Eq. (\ref{IN}), the spectra $S_{I_{\pm}(0)}(\omega)$ of the currents $I_{\pm}(0)$ from Eq. (\ref{Ipm}) are found. Inserting $S_{I_{\pm}(0)}(\omega)$ into Eq. (\ref{HFDT}), the energy density at $x=0$ becomes
\begin{equation}
H'(0)=\frac{1}{2}L'\int_{-\infty}^{\infty}\frac{d\omega}{2\pi}\frac{\hbar \omega}{1-e^{-\beta\hbar\omega}}\frac{\frac{|t_1|^2}{R_1}(1+|r_2|^2)+\frac{|t_2|^2}{R_2}(1+|r_1|^2)}{2|1-r_1r_2e^{i2kl}|^2}.
\label{H0a}
\end{equation}
Using $1=|r_n|^2+(Z_0/R_n)|t_n|^2$ for $n=1,2$ (Appendix A), $k=\omega/c$, and the expressions for $c$ and $Z_0$ [Eqs. (\ref{HELM}) and (\ref{r})], we obtain
\begin{equation}
H'(0^+)=H'(0)=\int_0^{\infty}\frac{dk}{2\pi}\frac{\hbar ck}{\tanh\left(\frac{\beta\hbar ck}{2}\right)}
\frac{1-|r_1|^2|r_2|^2}{|1-r_1r_2e^{i2kl}|^2}.
\label{H0}
\end{equation}
In order to convert the integral to positive $k$-values we have used the fact that the absolute values of the linear responses $r_{1,2}$ and $1-r_1r_2e^{i2kl}$ are even functions of $k$, since they result from Fourier transforms of real functions. The above result is identical to $H'(0^+)=H'^{(C)}$ calculated within the QED formalism, Eq. (\ref{HC}). The physical picture provided by the QED scattering problem was that of non-absorptive mirrors and fields transmitted to infinity, equivalent to the specific case $R_1=R_2=Z_0$ in Fig. 4. The circuit problem of Fig. 4 is however more general, so that the above expression for $H'(0^+)$ is valid for reflection coefficients of \emph{any} complex impedances that effectively \emph{terminate} the line.

Since the FDT result for $H'(0^+)$, Eq. (\ref{H0}), is identical to that from  Eq. (\ref{HC}), it can also be renormalized in the same way, i.e. by subtraction of the free-TL result of Eq. (\ref{H1}). Whereas in the QED approach the physical justification for this subtraction could be attributed to a counter force acting at $x=0^-$ [Eq. (\ref{f})], a more constructive point-of-view is the latter Lifshitz-theory-like interpretation given above; namely, the stress $H'(0^+)$ at $x=0^+$ is renormalized by the subtraction of that of a free TL with the same line parameters $C'$ and $L'$ as in $x=0^+$.

To conclude, the TL circuit-mediated Casimir force between two general complex impedances $Z_{1,2}$ which terminate a TL with characteristic impedance $Z_0$ is given by
 \begin{equation}
f=\int_0^{\infty}\frac{dk}{2\pi}\frac{-\hbar ck}{\tanh\left(\frac{\beta\hbar ck}{2}\right)}\left[\frac{r_1r_2e^{i2kl}}{1-r_1r_2e^{i2kl}}+\frac{r_1^{\ast}r_2^{\ast}e^{-i2kl}}{1-r_1^{\ast}r_2^{\ast}e^{-i2kl}}\right],
\label{fFDT}
\end{equation}
with $r_{1,2}$ the reflection coefficients from Eq. (\ref{r}). This result is in contrast to that obtained by the QED approach, Eq. (\ref{fQED}), where the reflection coefficients are those from Eq. (\ref{rt}). Therefore, the above FDT-obtained result (Eq. \ref{fFDT}) extends its validity of its QED-obtained counterpart (Eq. \ref{fQED}) to include dissipative scatterers/impedances that terminate the line.

\section{Examples and sign of force}
We shall now illustrate the application of the general formula, Eq. (\ref{fFDT}), with a few simple examples at zero temperature, where we take the two impedances to terminate the TL (as in Fig. 4). In particular, we wish to focus on the sign of the force, either attractive or repulsive, which is facilitated by performing the integral (\ref{fFDT}) in imaginary frequencies, the so-called Wick rotation \cite{FoQVe}. Out of the four examples considered below, the first two are of pedagogical nature, whereas the latter two may also have practical implications.

\subsection{Rotation to imaginary frequency}
We begin by rewriting Eq. (\ref{fFDT}) at zero temperature ($\beta\rightarrow \infty$) as
 \begin{equation}
f=-\frac{\hbar c}{\pi l^2}\mathrm{Re}[I], \quad I=\int_0^{\infty}dv v\frac{r_1(v)r_2(v)e^{i2v}}{1-r_1(v)r_2(v)e^{i2v}},
\label{I}
\end{equation}
with $v=kl=\omega(l/c)$ being a dimensionless frequency. Considering that the reflections $r_{1,2}$ and the Fabry-P\'{e}rot cavity response $1-r_1r_2e^{i2kl}$ are causal linear response functions, they are analytic in the upper half complex plane as a function of $v$ (dimensionless frequency $\omega$, extended to be a complex variable) \cite{LL} (this means that any pole/resonance that appears on the real axis is slightly shifted below the real axis, i.e. includes dissipation, to account for realistic causal systems). Therefore, the integrand of $I$ in (\ref{I}) must also be analytic in the upper half plane, where it vanishes exponentially for $|v|\rightarrow\infty$. This allows to deform the contour of integration from the real axis of $v$ into the imaginary axis \cite{FoQVe}. The resulting integral in terms of the dimensionless imaginary frequency $u=-iv$ becomes
 \begin{equation}
f=\frac{\hbar c}{\pi l^2}\int_0^{\infty}du u \frac{r_1(iu)r_2(iu)e^{-2u}}{1-r_1(iu)r_2(iu)e^{-2u}},
\label{fu}
\end{equation}
where for the evaluation of $r_{1,2}(iu)$ we recall that $iu=\omega(l/c)$. We also note that this integral is guaranteed to take real values since linear response functions are real and monotonically decreasing in the positive imaginary axis \cite{LL}.

This also means that $r_{1,2}(iu)$ take only real values smaller than $r_{1,2}(i0)<1$, which is extremely useful in determining the sign of the force. The denominator must therefore be positive such that the sign of the integrand is exclusively determined by the product $r_{1}(iu)r_{2}(iu)$ in the numerator. Considering impedances $Z_{1,2}$ that terminate the line, with the reflection coefficients of Eq. (\ref{r}), repulsion ($f<0$) is then obtained (per frequency) for
\begin{equation}
Z_1(iu)<Z_0<Z_2(iu) \Rightarrow r_{1}(iu)r_{2}(iu)<0,
\label{rep}
\end{equation}
where $1$ and $2$ are interchangeable. This is in analogy to the known condition in 3d \cite{DZO,CAP}, which is nevertheless typically stated in terms of electric permittivity instead of impedance, hence ignoring the magnetic response.

\subsection{Example 1: a pair of short/open circuits}
Here we consider that both impedances are either shorted, i.e. replaced by a wire so that $Z_{1,2}=0$, or disconnected $Z_{1,2}=\infty$ ("open circuit", equivalent to vanishing capacitance or infinite inductance). Whereas the first case gives $r_{1,2}=-1$ and the latter $r_{1,2}=1$ [Eq. (\ref{r})], both cases yield $r_{1}r_{2}=1$ at all frequencies $u$, and hence to an attractive force. This case constitutes the 1d analogue to the original situation considered by Casimir \cite{CAS}. Performing the integral in Eq. (\ref{fu}) with $r_{1}r_{2}=1$ we obtain $f=\pi\hbar c/(24 l^2)$ as expected in 1d \cite{REY,BOY1}.

\subsection{Example 2: open circuit in front of short circuit}
Consider now a short circuit, $Z_1=0$, interacting with an open circuit, $Z_2=\infty$, satisfying the condition (\ref{rep}) for a repulsive force. Inserting $r_{1}r_{2}=-1$ into Eq. (\ref{fu}) for all frequencies $u$, we find $f=-\pi\hbar c/(48 l^2)$. This result was obtained in Ref. \cite{BOY1} by similar considerations to those of the original Casimir treatment, adopted to 1d field quantization with unlike boundary conditions. Here however, a concrete physical system of impedances inside circuits is considered, which allows an interpretation in analogy to the result by Boyer in Ref. \cite{BOY}, where the free-space Casimir force between a perfect magnetic conductor $\mu(iu)\rightarrow\infty$ and a perfect electric conductor $\epsilon(iu)\rightarrow\infty$ was found to be repulsive.
Therefore, the impedance $Z_1=0$ can be viewed as that of a capacitor $Z_1(iu)=l/(cCu)$ in the limit $C\rightarrow \infty$ in analogy to $\epsilon(iu)\rightarrow\infty$, whereas $Z_2=\infty$ can be seen as an inductor $Z_2(iu)=cLu/l$ in the limit of infinite inductance analogous to $\mu(iu)\rightarrow\infty$.

\begin{figure}
\begin{center}
\includegraphics[width=\columnwidth]{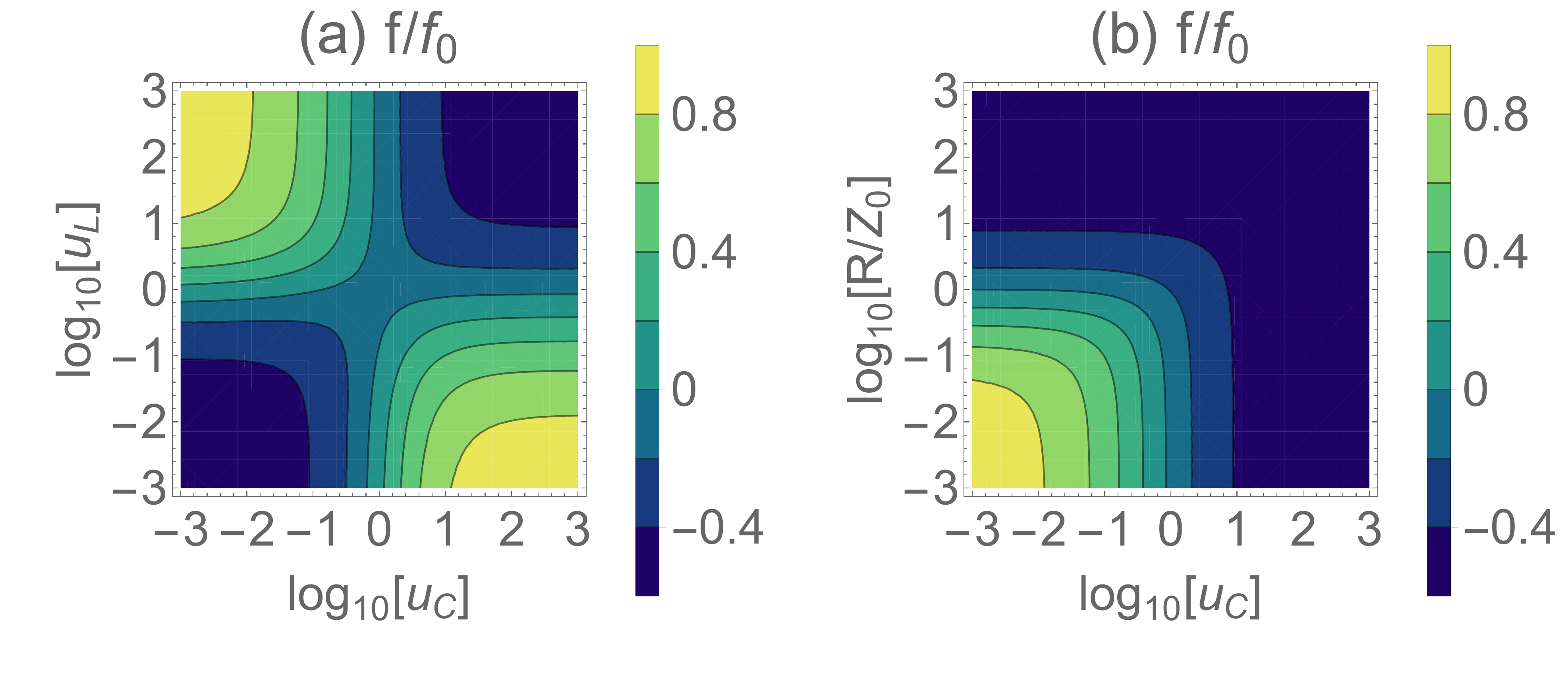}
\caption{\small{
Force between circuit elements which terminate the TL: Examples. (a) Force between a capacitor $C$ and an inductor $L$ (example 3) as a function of their respective parameters $u_C$ and $u_L$ from Eq. (\ref{rCrL}). The force is in units of the attractive force $f_0=\pi\hbar c/(24 l^2)$ between two identical perfect 1d mirrors. Both attractive ($f/f_0>0$) and repulsive ($f/f_0<0$) forces can be obtained, with the asymptotic limiting values of $f/f_0=1$ and $f/f_0=-1/2$ from examples 1 and 2, respectively. (b) Same as (a) for the first circuit element being a capacitor $C$ in series with a resistor $R$, and the second element being a short circuit ($Z_2=0$).
 }} \label{fig5}
\end{center}
\end{figure}

\subsection{Example 3: capacitor in front of inductor}
The former two examples can be made more realistic by considering circuit elements such as a capacitor $C$ and an inductor $L$. The frequency dependencies of their respective impedances, $Z_1=i/(\omega C)$ and $Z_2=-i\omega L$, exhibit opposite limiting behaviors at $\omega\rightarrow0$ and $\omega\rightarrow\infty$, namely, when $Z_1$ appears as open circuit $Z_2$ appears as short circuit and vise versa. Therefore, both behaviors found in examples 1 and 2 above should appear in this case. More explicitly, consider their corresponding reflection coefficients in imaginary frequency
\begin{eqnarray}
&&r_1(iu)=\frac{u_C-u}{u_C+u}, \quad u_C=\frac{l}{c}\frac{1}{Z_0 C},
\nonumber\\
&&r_2(iu)=\frac{u-u_L}{u+u_L}, \quad u_L=\frac{l}{c}\frac{Z_0}{L},
\label{rCrL}
\end{eqnarray}
where $u_C$ and $u_L$ are the scaled frequencies associated with the capacitor and inductor, respectively. Figure 5b displays the force between the components, which is calculated by numerically integrating Eq. (\ref{fu}) as a function of the parameters $u_C$ and $u_L$. Indeed, both of the results from examples 1 (attractive force) and 2 (repulsive force) emerge at the following limits: when the capacitance is small ($u_C\rightarrow \infty$) $Z_1$ acts as open circuit and vise versa, whereas when the inductance is small ($u_L\rightarrow \infty$) $Z_2$ acts as short circuit and vise versa.

\subsection{Example 4: capacitor + resistor in front of short circuit}
All previous examples included non-dissipative impedances. We shall now exploit the general formula, Eq. (\ref{fFDT}), derived using the FDT approach to consider a dissipative case that goes beyond that treated using the QED approach. Specifically, we consider an impedance $Z_1=i/(\omega C)+R$ of a capacitor in series with a resistor. The corresponding reflection coefficient in imaginary frequency is given by
\begin{equation}
r_1(iu)=\frac{(R/Z_0-1)u+u_C}{(R/Z_0+1)u+u_C},
\label{rRC}
\end{equation}
which changes sign as a function of $u$. Figure 5c presents the resulting force for $Z_2=0$ (short circuit), plotted as a function of the parameters $u_C$ and $R/Z_0$. Again, for small capacitance, where $u_C\gg 1$, the capacitor, and hence $Z_1$, behave as open circuit, so that the repulsion of example 2 is recovered.

\section{Discussion}
We have considered the fluctuation force between two neutral objects connected by wires. The objects are characterized by their impedances and, within the FDT approach, the fluctuations emerge from their dissipative part. This approach is simple to generalize for any dissipative circuit, and in particular to lossy TLs, by modelling any resistance with a current source as in Fig. 2 \cite{HEN}. Moreover, it is also natural to apply to any general circuit configuration such as lumped circuits, since it does not rely on the quantization of electromagnetic waves, as in the QED approach. As such, it can become useful in the exploration of a variety of fluctuation phenomena in circuits, such as those studied in Refs. \cite{CAScir,WOOD}, by merely solving a set of Kirchhoff-law equations. In this respect, the study of fluctuation forces in circuits could provide a conceptually simple system wherein Casimir physics can be explored in the absence of technical complications which may not add to its essence.

We note that in principal, the general configuration beyond that of Fig. 1c can also be calculated within the QED scattering formalism. That would require to replace any resistive element $R$ by a semi-infinite TL with characteristic impedance $R$ and corresponding quantum field fluctuations. This would result in a more complicated scattering problem with multiple channels and inputs which is nevertheless fully equivalent to the more elegant circuit solution of the FDT formalism. In this respect, let us also refer back to the fact that the result for the force in the QED approach, Eq. (\ref{fQED}), is identical to that found using the FDT approach, Eq. (\ref{fFDT}), even though the latter was derived for the general case whereas the former for the specific case of non-dissipative impedances embedded in a uniform TL. In fact, a similar situation exists also in 3d derivations of the Casimir force between dielectric slabs or mirrors. There, the Lifshitz treatment, which is performed for generally lossy dielectrics \cite{DZO}, formally leads to the same result obtained by derivations which assume lossless dielectrics and mirrors \cite{REY,MIL}.

Although the main focus of this paper is on formalism, we shall conclude with a brief account on possible experimental realizations. A straightforward realization where the force between the circuit elements would in principle be observable is a systems wherein the motion of the center of mass of one of the elements is allowed and is measurable. A different approach would be to consider the  interaction potential built between the two elements, which is associated with the force. Then, if the energy spectrum describing the internal degrees of freedom of one of the elements is quantized, such as in a super-conducting qubit, the interaction potential could possibly be measured via the shift it induces on these energy levels \cite{vdWTL,WAL4,NAK}. Alternatively, in case one of the interacting elements possesses a movable internal coordinate, such as in a variable capacitor or inductor, the interaction potential may be inferred from the force induced on this coordinate \cite{CAScir,WOOD,TEU}.

\begin{acknowledgements}
I would like to thank Ulf Leonhardt for valuable discussions and Itay Griniasty for careful reading of the manuscript. The financial support of the European Research Council (ERC), the Israel Science Foundation (NSF) and the MIT-Harvard Center for Ultracold Atoms is acknowledged.
\end{acknowledgements}

\appendix
\section{Relation between reflection and transmission coefficients}
By energy conservation, it is clear that the relation between the power reflection and transmission coefficients of a non-dissipative 1d scatterer satisfies $|r|^2+|t|^2=1$. Similar energy conservation reasoning lead to the modification of this relation when the scatterer is situated at the boundary between media with different impedances. We will now show how these relations can be obtained using simple circuit analysis for the reflection of an impedance $Z$ that \emph{terminates} a TL, Eq. (\ref{r}), when the transmission (absorption) is defined as $t\equiv1+r=2Z/(Z+Z_0)$.
Denoting $Z'=\mathrm{Re}[Z]$ and $Z''=\mathrm{Im}[Z]$, we obtain
\begin{eqnarray}
&&1-|r|^2=\frac{4Z_0^2\left(|Z|^2Z'/Z_0+2Z'^2+Z'Z_0\right)}{\left[|z|^2+Z_0(2Z'+Z_0)\right]^2},
\nonumber\\
&&|t|^2=\frac{4|Z|^2\left(|Z|^2+2Z_0Z'+Z_0^2\right)}{\left[|z|^2+Z_0(2Z'+Z_0)\right]^2}.
\label{A1}
\end{eqnarray}
Then, representing the impedance by a parallel resistor $R$ and reactance $X$ as in Fig. 4b, we insert the relations (\ref{Zn}) into Eqs. (\ref{A1}) and obtain
\begin{equation}
1=|r|^2+\frac{Z_0}{R}|t|^2.
\label{A2}
\end{equation}
The relation to energy conservation in 1d scattering becomes clear if we think of the resistor $R$ in Fig. 4b as a semi-infinite TL extending to left side, with a characteristic impedance $R$. Then, Eq. (\ref{A2}) describes the energy transfer between a medium with impedance $R$ to a medium with impedance $Z_0$ through a lossless scatterer $i X$, and therefore the ratio of impedances appears. For a lossless mirror between identical media, $R=Z_0$, we obtain the usual relation  $|r|^2+|t|^2=1$.

\section{Force exerted on circuit components}
Consider a lumped-circuit component embedded in a TL at position $x_0$, as in Fig. 6. The force acting on the component is that exerted by the electromagnetic fields on a volume $V$ which surrounds the component (blue dashed line in Fig. 6) \cite{FoQVs,NH},
\begin{equation}
\mathbf{f}=\int_{\partial V} d\mathbf{a} \cdot \overline{\overline{T}}(\mathbf{r}).
\label{ff}
\end{equation}
Here $d\mathbf{a}$ is a vector of an area element directed normal to the surface $\partial V$ confining the volume $V$, and $\overline{\overline{T}}(\mathbf{r})$ is the Maxwell stress tensor (outside of the component)
\begin{equation}
T_{ij}=\epsilon_0E_iE_j+\mu_0H_iH_j-W\delta_{ij}, \quad W=\frac{1}{2}\left(\epsilon_0 \mathbf{E}^2+\mu_0 \mathbf{B}^2\right).
\label{T}
\end{equation}
Since we are interested in forces mediated (exerted) by the TEM field mode of the TL, we consider $T_{ij}$ for electric and magnetic fields perpendicular to each other and to the propagation axis $x$, $E_i=E \delta_{iy}$ and $H_i=H \delta_{iz}$, yielding a diagonal stress tensor $T_{ij}=T_{ii}\delta_{ij}$ with $T_{xx}=-W$. The force along the positive $x$ direction, $f=\mathbf{e}_x\cdot\mathbf{f}=\int_{\partial V}da_xT_{xx}$ is then found to be
\begin{eqnarray}
f&=&-\iint dydz T_{xx}(x_0^-,y,z)+\iint dydz T_{xx}(x_0^+,y,z)
\nonumber\\
&=&\iint dydz \left[W(x_0^-,y,z)- W(x_0^+,y,z)\right]
\nonumber\\
&=&H'(x_0^-)-H'(x_0^+).
\label{fff}
\end{eqnarray}
Here we used the fact that the integral of the electromagnetic energy density $W(x,y,z)$ over $y$ and $z$ is equivalent to the energy per unit length $H'(x)$.

Considering now the impedance 1 from the Casimir configuration of Fig. 1c, the force in the positive $x$ direction, $f_1=H'(0-)-H'(0^+)$, is the attractive force acting on this circuit element. For the impedance 2, the attractive force is that in the negative $x$ direction and is hence given by $f_2=H'(l^+)-H'(l^-)$. Finally, since the energy density is constant at each section of the TL [see Eq. (\ref{HQED})], the two forces are identical as expected, $f_1=f_2\equiv f$.
\begin{figure}
\begin{center}
\includegraphics[scale=0.4]{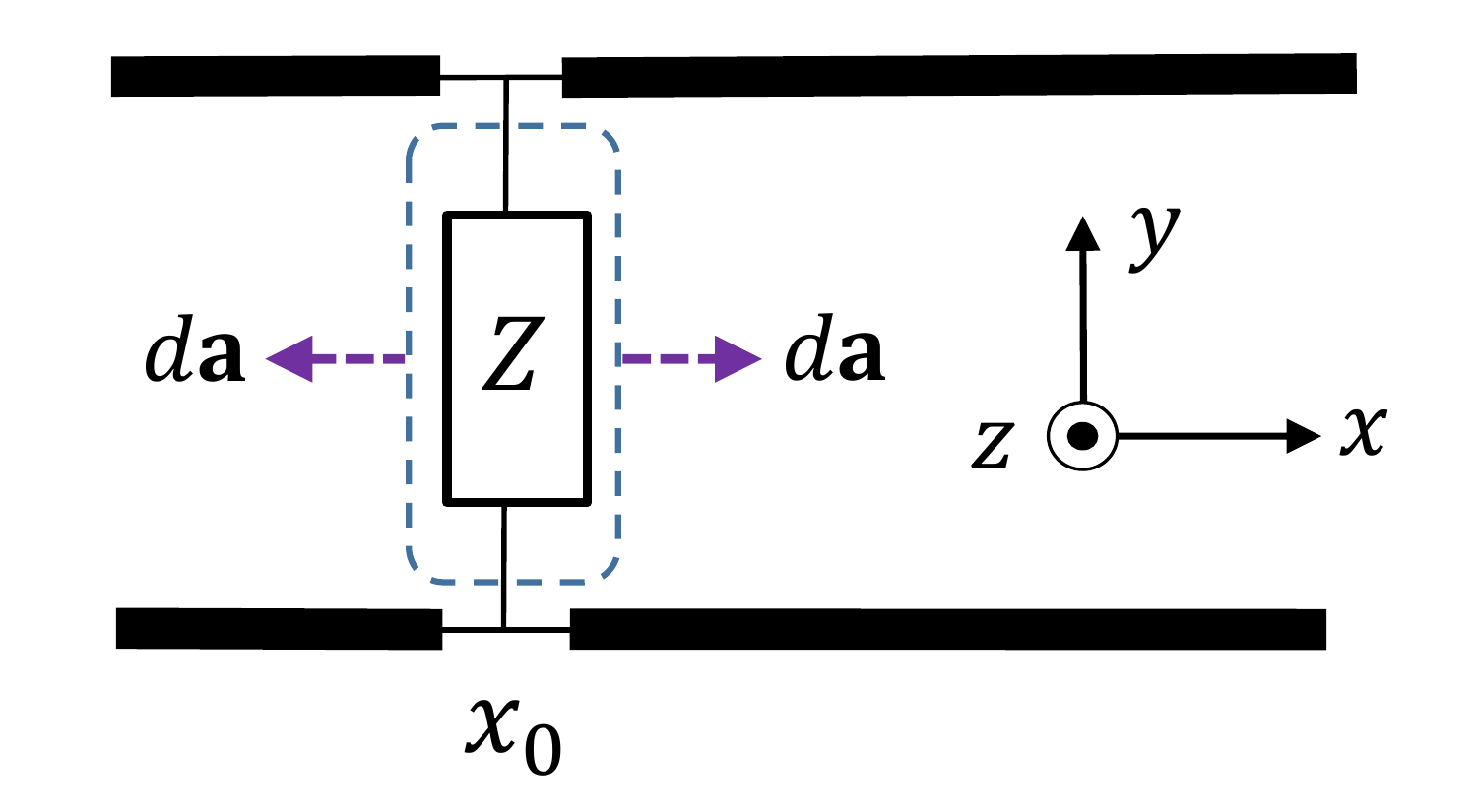}
\caption{\small{
The force exerted on a circuit component: see text in Appendix B.
 }} \label{fig6}
\end{center}
\end{figure}

\section{Quantization of the homogenous transmission line}
In the following, we provide a brief account of the quantization of the TL. The voltage, current, and hence the charge fields in a homogenous TL satisfy the 1d Helmholtz equation, Eq. (\ref{HELM}). It can be easily shown that this equation, e.g. for the charge $Q(x,t)$, can be derived from the following Lagrangian,
\begin{equation}
\mathcal{L}=\int dx \mathcal{L}'(x)=\int dx\frac{1}{2}\left[L'(\partial_t Q)^2-\frac{1}{C'}(\partial_x Q)^2\right].
\label{B1}
\end{equation}
The conjugate variable to $Q(x)$ is found to be the flux per unit length, $\phi(x)\equiv\partial\mathcal{L}'/\partial \dot{Q}(x)=L'I(x)$, resulting in the Hamiltonian,
\begin{equation}
H=\int dx\frac{1}{2}\left[\frac{1}{L'}\phi^2+\frac{1}{C'}(\partial_x Q)^2\right].
\label{B2}
\end{equation}
Canonical quantization is performed by demanding the commutation relations $[\hat{Q}(x),\hat{Q}(x')]=0$, $[\hat{\phi}(x),\hat{\phi}(x')]=0$ and $[\hat{Q}(x),\hat{\phi}(x')]=i\hbar \delta(x-x')$. Using the expansion $\hat{Q}(x)=\sum_k Q_k \hat{a}_k e^{ikx}/\sqrt{L}+\mathrm{h.c.}$ and hence $\hat{\phi}(x)=-\sum_k ic|k|L'Q_k \hat{a}_k e^{ikx}/\sqrt{L}+\mathrm{h.c.}$ with $[\hat{a}_k,\hat{a}_{k'}]=0$ and $[\hat{a}_k,\hat{a}_{k'}^{\dag}]=\delta_{kk'}$, we find the coefficients $Q_k=\sqrt{\hbar/(2Z_0|k|)}$. Finally, by using the telegraphers equation $\hat{V}(x)=-(1/C')d\hat{Q}/dx$ and the definition of the flux  $\hat{\phi}(x)=L'\hat{I}(x)$, we find the voltage and current operators from Eqs. (\ref{VI}) and (\ref{a}).

\section{Current fluctuations in resistors}
A simple way to derive Eq. (\ref{IN}) goes as follows. Treating the charge $Q$ on a lumped circuit element as its relevant dynamical variable, and considering an applied voltage $V(t)$, we obtain the interaction Hamiltonian in a linear form $H=QV(t)$. Considering the impedance of the circuit element $Z(\omega)$, and using $I(\omega)=V(\omega)/Z(\omega)$ and $\dot{Q}(t)=I(t)$, we find $Q(\omega)=\chi(\omega)V(\omega)$ with the linear response $\chi(\omega)=i/\omega Z(\omega)$. Then, inserting $\mathrm{Im}[\chi]=i\mathrm{Re}[Z]/(\omega|Z|^2)$ into the FDT \cite{LL} we find the charge fluctuations
\begin{equation}
S_Q(\omega)=2\hbar \frac{\mathrm{Re}[Z]}{\omega |Z|^2}\frac{1}{1-e^{-\beta\hbar\omega}}.
\label{C1}
\end{equation}
Finally, the current fluctuations are obtained from $S_I(\omega)=\omega^2S_Q(\omega)$, where Eq. (\ref{IN}) corresponds to the case $Z=R$.


\begin{thebibliography}{}
\bibitem{RAU1} E. Vetsch, D. Reitz, G. Sagu\'{e}, R. Schmidt, S. T. Dawkins and A. Rauschenbeutel, Phys. Rev. Lett. \textbf{104}, 203603 (2010).
\bibitem{RAU3} C. Sayrin, C. Junge, R. Mitsch, B. Albrecht, D. O'Shea, P. Schneeweiss, J. Volz and A. Rauschenbeutel, Phys. Rev. X \textbf{5}, 041036 (2015).
\bibitem{KIM2} A. Goban, C.-L. Hung, S.-P. Yu,	J. D. Hood,	J. A. Muniz, J. H. Lee,	M. J. Martin,	A. C. McClung,	K. S. Choi,	D. E. Chang,	 O. Painter	 and H. J. Kimble, Nat. Commun. \textbf{5}, 3808 (2014).	
\bibitem{KIM3} A. Goban, C.-L. Hung, J. D. Hood, S.-P. Yu, J. A. Muniz, O. Painter and H. J. Kimble, Phys. Rev. Lett. \textbf{115}, 063601 (2015).
%\bibitem{TOM} J. D. Thompson, T. G. Tiecke, N. P. de Leon, J. Feist, A. V. Akimov, M. Gullans, A. S. Zibrov, V. Vuleti\'{c} and M. D. Lukin, Science \textbf{340}, 1202 (2013).
\bibitem{LOD} M. Arcari, I. S\"{o}llner, A. Javadi, S. Lindskov Hansen, S. Mahmoodian, J. Liu, H. Thyrrestrup, E. H. Lee, J. D. Song, S. Stobbe, and P. Lodahl, Phys. Rev. Lett. \textbf{113}, 093603 (2014).
\bibitem{ALP} A. Sipahigil, R. E. Evans, D. D. Sukachev, M. J. Burek, J. Borregaard, M. K. Bhaskar, C. T. Nguyen, J. L. Pacheco, H. A. Atikian, C. Meuwly, R. M. Camacho, F. Jelezko, E. Bielejec, H. Park, M. Lon\u{c}ar, M. D. Lukin, Science \textbf{354}, 847 (2016).
%\bibitem{ORZ} J. A. Grover, P. Solano, L. A. Orozco, and S. L. Rolston, Phys. Rev. A \textbf{92}, 013850 (2015).
\bibitem{WAL1} A. F. van Loo, A. Fedorov, K. Lalumi\`{e}re, B. C. Sanders, A. Blais, and A. Wallraff, Science \textbf{342}, 1494 (2013).
\bibitem{WAL2} J. A. Mlynek, A. A. Abdumalikov,	C. Eichler and A. Wallraff, Nat. Commun. \textbf{5}, 5186 (2014).
\bibitem{GON} A. Gonzalez-Tudela, D. Martin-Cano, E. Moreno, L. Martin-Moreno, C. Tejedor and F. J. Garcia-Vidal, Phys. Rev. Lett \textbf{106}, 020501 (2011).
\bibitem{RDDI} E. Shahmoon and G. Kurizki, Phys. Rev. A \textbf{87}, 033831 (2013).
\bibitem{CHA1} D. E. Chang, J. I. Cirac and H. J. Kimble, Phys. Rev. Lett. \textbf{110}, 113606 (2013).
\bibitem{RIT} T. Grie{\ss}er and H. Ritsch, Phys. Rev. Lett. \textbf{111}, 055702 (2013).
\bibitem{GOR} Z. Eldredge, P. Solano, D. Chang and A. V. Gorshkov, Phys. Rev. A \textbf{94}, 053855 (2016).
\bibitem{RAB} G. Calaj\'{o}, F. Ciccarello, D. Chang and P. Rabl, Phys. Rev. A \textbf{93}, 033833 (2016).
\bibitem{LIDDIna} E. Shahmoon, I. Mazets and G. Kurizki, Opt. Lett. \textbf{39}, 3674 (2014).
\bibitem{CHA2} J. S. Douglas, H. Habibian, C.-L. Hung, A. V. Gorshkov, H. J. Kimble and D. E. Chang, Nat. Photon. \textbf{9}, 326 (2015).
\bibitem{EITNLO} E. Shahmoon, P. Grisins, H. P. Stimming, I. Mazets and G. Kurizki, Optica \textbf{3}, 725 (2016).
\bibitem{CHA3} J. S. Douglas, T. Caneva and D. E. Chang, Phys. Rev. X \textbf{6}, 031017 (2016).
\bibitem{vdWTL} E. Shahmoon, I. Mazets and G. Kurizki, Proc. Natl. Acad. Sci. USA \textbf{111}, 10485 (2014).
\bibitem{MAZ} E. \'{A}lvarez and F. D. Mazzitelli, Phys. Rev. D \textbf{79}, 045019 (2009).
\bibitem{SIL} S. I. Maslovski and M. G. Silveirinha, Phys. Rev. A \textbf{82}, (2010).
\bibitem{vdWMWG} E. Shahmoon and G. Kurizki, Phys. Rev. A  \textbf{87}, 062105 (2013).
\bibitem{HAK} H. R. Haakh and S. Scheel, Phys. Rev. A \textbf{91}, 052707 (2015).
\bibitem{FAR} R. de Melo e Souza, W. J. M. Kort-Kamp, F. S. S. Rosa and C. Farina, Phys. Rev. A \textbf{91}, 052708 (2015).
\bibitem{SCH} S. Scheel, S. Y. Buhmann, C. Clausen, and P. Schneeweiss, Phys. Rev. A \textbf{92}, 043819 (2015).
\bibitem{WAL3} C. Eichler, D. Bozyigit, C. Lang, M. Baur, L. Steffen, J. M. Fink, S. Filipp, and A. Wallraff, Phys. Rev. Lett. \textbf{107}, 113601 (2011).
\bibitem{SID} K. W. Murch, S. J. Weber, K. M. Beck, E. Ginossar, I. Siddiqi, Nature, \textbf{499}, 62 (2013).
\bibitem{WIL} C. M. Wilson, G. Johansson, A. Pourkabirian, M. Simoen, J. R. Johansson, T. Duty, F. Nori and P. Delsing, Nature \textbf{479}, 376-379 (2011).
\bibitem{PARA} P. L\"{a}hteenm\"{a}ki, G. S. Paraoanu, J. Hassel and J. Hakonen,  Proc. Natl. Acad. Sci. \textbf{110}, 4234 (2013).
\bibitem{YUR} B. Yurke and J. S. Denker, Phys. Rev. A \textbf{29}, 1419 (1984).
\bibitem{DEV} M. H. Devoret, \emph{Quantum Fluctuations in Electrical Circuits}, in \emph{Quantum Fluctuations}, Les Houches Session LXIII, pp. 351 (Elsevier 1997).
\bibitem{REY} M. T. Jaekel and S. Reynaud, Journal de Physique I \textbf{1}, 1395 (1991).
\bibitem{LL} L. D. Landau and E. M. Lifshitz, \emph{Statistical Physics, Part 1} (Pergamon Press, 3rd Edtion 1980).
\bibitem{DZO} I. E. Dzyaloshinskii , E. M. Lifshitz and L. P. Pitaevskii, Advances in Physics, \textbf{10}, 165 (1961).
\bibitem{BAR} Y. S. Barash and V. L. Ginzburg, Sov. Phys. Usp. \textbf{27}, 467 (1984).
\bibitem{MIL} P. W. Milonni, \emph{The Quantum Vacuum: An Introduction to Quantum Electrodynamics} (Academic, 1993).
\bibitem{FoQVs} S. Scheel, in \emph{Forces of the Quantum Vacuum} (W. M .R. Simpson and U. Leonhardt, Eds.), Ch. 3 (World Scientific, 2015).
\bibitem{KONG} J. A. Kong, \emph{Electromagnetic Wave Theory}, (John Wiley and Sons, Inc., 1986).
\bibitem{POZ} D. M. Pozar \emph{Microwave Engineering} (John Wiley and Sons, 2005).
\bibitem{BLA} A. Blais, R. S. Huang, A. Wallraff, S. M. Girvin and R. J. Schoelkopf, hys. Rev. A \textbf{69}, 062320 (2004).
\bibitem{HEN} J. R. Zurita-S\'{a}nchez and C. Henkel, Phys. Rev. A \textbf{73}, 063825 (2006).
\bibitem{NYQ} H. Nyquist, Phys. Rev. \textbf{32}, 110 (1928).
\bibitem{FoQVe} E. Shahmoon, in \emph{Forces of the Quantum Vacuum} (W. M .R. Simpson and U. Leonhardt, Eds.), Ch. 3 (World Scientific, 2015).
\bibitem{CAP} J. Munday, F. Capasso ans V. A. Parsegian, Nature \textbf{457}, 170 (2009).
\bibitem{CAS} H. B. G. Casimir, Proc. K. Ned. Akad. Wet. \textbf{51}, 793 (1948).
\bibitem{BOY1} T. H. Boyer, Am. J. Phys. \textbf{71}, 990 (2003).
\bibitem{BOY} T. H. Boyer, Phys. Rev. A \textbf{9}, 2078 (1974).
\bibitem{CAScir} E. Shahmoon and U. Leonhardt, arXiv:1612.03250.
\bibitem{WOOD} D. Drosdoff, I. V. Bondarev, A. Widom, R. Podgornik and L. M. Woods, Phys. Rev. X \textbf{6}, 011004 (2016).
\bibitem{WAL4} A. Fragner, M. G\"{o}ppl, J. M. Fink, M. Baur, R. Bianchetti, P. J. Leek, A. Blais and A. Wallraff, Science \textbf{322}, 1357 (2008).
\bibitem{NAK} Y. Tabuchi, S. Ishino, A. Noguchi, T. Ishikawa, R. Yamazaki, K. Usami and Y. Nakamura, Science \textbf{349}, 405 (2015).
\bibitem{TEU} J. D. Teufel, T. Donner, Dale Li, J. W. Harlow, M. S. Allman, K. Cicak, A. J. Sirois, J. D. Whittaker, K. W. Lehnert and R. W. Simmonds, Nature \textbf{475}, 359 (2011).
\bibitem{NH} L. Novotny and B. Hecht,\emph{ Principles of Nano-Optics} (Cambridge University Press, 2006).






\end{thebibliography}
\end{document}